\documentclass[conference, a4paper]{IEEEtran}
\IEEEoverridecommandlockouts
\usepackage[font=footnotesize,caption=false]{subfig} % preload with correct options...

\usepackage{cite}
\usepackage{amsmath,amssymb,amsfonts}
\usepackage{algorithmic}
\usepackage{graphicx}
\usepackage{textcomp}
\usepackage{xcolor}
\usepackage{booktabs}
\usepackage{glossaries}
\usepackage{subfig}
\usepackage{algorithm}
\usepackage{algorithmic}
\usepackage{bbm}
\usepackage{bm}
\usepackage{multirow}
\usepackage{tabularx}
\usepackage[utf8]{inputenc}
\usepackage{tikz}
\usetikzlibrary{shapes,arrows}
\usetikzlibrary{plotmarks}
\usetikzlibrary{patterns}
\usetikzlibrary{arrows.meta}
\usetikzlibrary{positioning}
\usetikzlibrary{spy}

\usepackage{pgfplots}
\usepgfplotslibrary{colorbrewer}
\usepgfplotslibrary{fillbetween}

\def\BibTeX{{\rm B\kern-.05em{\sc i\kern-.025em b}\kern-.08em
    T\kern-.1667em\lower.7ex\hbox{E}\kern-.125emX}}
\newacronym{fec}{FEC}{forward error correction}
\newacronym{3gpp}{3GPP}{3rd Generation Partnership Project}
\newacronym{6g}{6G}{the sixth-generation of mobile systems}    
\newacronym{iot}{IoT}{Internet of Things}
\newacronym{ntn}{NTN}{Non-Terrestrial Network}
\newacronym{leo}{LEO}{Low Earth Orbit}
\newacronym{geo}{GEO}{Geosynchronous Earth Orbit}
\newacronym{isl}{ISL}{Inter-Satellite Link}
\newacronym{gsl}{GSL}{Ground-to-Satellite Link}
\newacronym{qos}{QoS}{Quality of Service}
\newacronym{ofdma}{OFDMA}{Orthogonal Frequency-Division Multiple Access}
\newacronym{b5g}{B5G}{5G and Beyond}
\newacronym{mimo}{MIMO}{Multiple-Input Multiple-Output}
\newacronym{embb}{eMBB}{Enhanced Mobile Broadband}
\newacronym{urllc}{URLLC}{ultra-reliable and low-latency communications}
\newacronym{mmtc}{mMTC}{massive machine-type communications}
\newacronym{ue}{UE}{User Equipment}
\newacronym{nr}{NR}{New Radio}
\newacronym{gap}{GAP}{Generalized Assignment Problem}
\newacronym{mgap}{MGAP}{Multi-Level Generalized Assignment Problem}
\newacronym{csi}{CSI}{channel state information}    
\newacronym{ran}{RAN}{radio access network}
\newacronym{5g}{5G}{the 5th generation of mobile networks}
\newacronym{uav}{UAV}{unmanned aerial vehicle}
\newacronym{ap}{AP}{access point}
\newacronym{sic}{SIC}{successive interference cancellation}
\newacronym{mmimo}{mMIMO}{massive \gls{mimo}}
\newacronym{snr}{SNR}{signal-to-noise ratio}
\newacronym{sinr}{SINR}{signal-to-interference plus noise ratio}
\newacronym{fifo}{FIFO}{first-in first-out}
\newacronym{mab}{MAB}{multi-armed bandit}
\newacronym{rl}{RL}{reinforcement learning}
\newacronym{noma}{NOMA}{non-orthogonal multiple access}
\newacronym{rv}{RV}{random variable}
\newacronym{drl}{DRL}{deep \gls{rl}}
\newacronym{irsa}{IRSA}{irregular repetition slotted ALOHA}
\newacronym{oma}{OMA}{orthogonal multiple access}
\newacronym{fdma}{FDMA}{frequency division multiple access}
\newacronym{tdma}{TDMA}{time division multiple access}
\newacronym{mdp}{MDP}{Markov decision process}
\newacronym{bs}{BS}{base station}
\newacronym{rsma}{RSMA}{rate-splitting multiple access}
\newacronym{dtmc}{DTMC}{discrete-time Markov chain}
\newacronym{cscg}{CSCG}{circularly symmetric complex Gaussian}
\newacronym{ack}{ACK}{acknowledgement}
\newacronym{nack}{NACK}{negative ACK}
\newacronym{aoi}{AoI}{age-of-information}
\newacronym{tare}{TRE}{time-averaged reconstruction error}
\newacronym{tacae}{TCAE}{time-averaged cost of actuation error}
\newacronym{udc}{UC}{update-delivery cost}

\begin{document}
\title{{Coexistence of Real-Time {Source Reconstruction} and Broadband Services Over Wireless Networks}}
%Flexible RAN sharing among heterogeneous services for process monitoring}}
%RAN slicing for Process Monitoring}
\author{\IEEEauthorblockN{Anup Mishra\IEEEauthorrefmark{1}, Nikolaos Pappas\IEEEauthorrefmark{2}, {\v C}edomir Stefanovi{\'c}\IEEEauthorrefmark{1}, Onur Ayan\IEEEauthorrefmark{3},
 Xueli An\IEEEauthorrefmark{3},\\ Yiqun Wu\IEEEauthorrefmark{3}, Petar Popovski\IEEEauthorrefmark{1}, and Israel Leyva-Mayorga\IEEEauthorrefmark{1}}\IEEEauthorblockA{\IEEEauthorrefmark{1}Department of Electronic systems, Aalborg University, Denmark (\{anmi,  cs, petarp, ilm\}@es.aau.dk)\\
 \IEEEauthorrefmark{2}Link\"oping University, Link\"oping, Sweden (nikolaos.pappas@liu.se)\\
  \IEEEauthorrefmark{3}Huawei Technologies, Munich, Germany (\{onur.ayan, xueli.an, wuyiqun\}@huawei.com)}\thanks{This work is a joint contribution by members of Working Item 205 on ``6G Radio Access'' of the one6G association.}\vspace{-0.5cm}}
\maketitle
\begin{abstract}
Achieving a flexible and efficient sharing of wireless resources among a wide range of novel applications and services is one of the major goals of \gls{6g}. Accordingly, this work investigates the performance of a real-time system that coexists with a broadband service in a frame-based wireless channel. Specifically, we consider real-time remote tracking of an information source, where a device monitors its evolution and sends updates to a \gls{bs}, which is responsible for real-time source reconstruction {and, potentially,} remote actuation. To achieve this, the \gls{bs} employs a grant-free access mechanism to serve the monitoring device together with a broadband user, which share the available wireless resources through orthogonal or non-orthogonal multiple access schemes. We analyse the performance of the system with time-averaged reconstruction error, time-averaged cost of actuation error, and update-delivery cost as performance metrics. Furthermore, we analyse the performance of the broadband user in terms of throughput and energy efficiency. Our results show that an orthogonal resource sharing between the users is beneficial in most cases where the broadband user requires maximum throughput. However,  sharing the resources in a non-orthogonal manner leads to a far greater energy efficiency.
\end{abstract}
\glsresetall
\vspace{-0.11cm}
\section{Introduction}\label{Intro_Section}
Real-time autonomous and cyber-physical systems have attracted significant attention over the past decade due to their broad range of time-critical applications such as  autonomous transportation and industrial automation~\cite{pappas@dtmcTWC,WEIMER2012178}. These systems have devices, often geographically distributed, monitoring an information source or a process and sending updates to a remote monitor and/or digital twin. The objective of such systems is real-time tracking and reconstruction of the source process at the monitor receiving the updates, which enables decision-making and actuation~\cite{pappas@dtmcTWC}. Several metrics, such as mean square estimation error, \gls{aoi} and its variants, etc., have been studied as performance measures for remote tracking\cite{WEIMER2012178,Sun@AoI}. These metrics, however, are oblivious to the semantics of information (i.e., significance, goal-oriented usefulness, and contextual
value) and its impact on the overall cyber-physical system\cite{TimePerspective@Popovski,Pappas@SmeanticsMagazine,pappas@dtmcTWC}. To address this problem, recent works introduced metrics such as real-time reconstruction error and cost of actuation error, which account for the semantics of information transmitted and capture different aspects of the system performance\cite{pappas@dtmcTWC,pappas@dtmc}.
\begin{figure}[t]
\centering
\includegraphics[width=0.64\columnwidth, height=3.9cm]{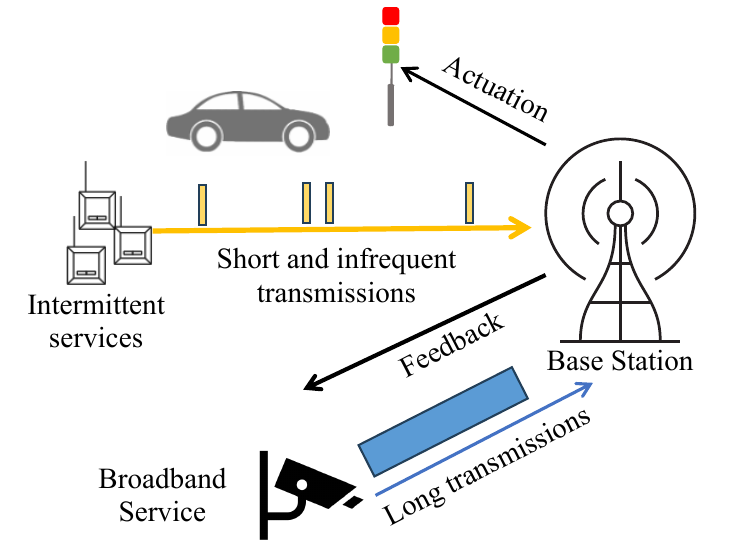}\vspace{-0.4cm}
\caption{A representative uplink scenario with a \gls{bs} serving an intermittent user and a broadband user.}\vspace{-0.60cm}
\label{fig:BBU_IU_Scenrio}\glsreset{bs}
\end{figure}
\par Studies on real-time remote tracking typically consider that updates from a device are followed by feedback from the monitor, in the form of an \gls{ack} or \gls{nack}. Moreover, these often rely on the `idealistic' assumptions of always having wireless resources available for transmitting updates and instantaneous feedback\cite{pappas@dtmcTWC}. However, such assumptions are unrealistic, as these would require an exorbitant wastage of radio resources, particularly given the intermittent nature of device transmissions in real-time autonomous systems\cite{Ley23Asilomar}. Specifically, major challenges in wireless communication systems include the long access delay and overhead in grant-based access mechanisms, the inefficient radio resource utilization in grant-free access mechanisms, and the delayed transmission of the update or feedback in frame-based transmissions \cite{Ley23Asilomar,OJCOMMS@fedrico}. The massive number of heterogeneous devices in future wireless networks, such as \gls{6g}, is going to exacerbate these challenges further. 
\textcolor{black}{Consequently, the performance of real-time remote reconstruction must be investigated in resource sharing communication scenarios that do not rely on idealistic assumptions. Moreover, the analysis must account for relevant performance metrics, the user distribution across service types, and the multiple access mechanisms that allocate resources to each service ~\cite{OJCOMMS@fedrico,Chiariotti2021,pappas@dtmc}.} 
\par Motivated by the above discussion, in this work we consider an uplink scenario where a device is monitoring an information source and sending updates to a \gls{bs}, which tracks and reconstructs the source process in real-time. {Fig.~\ref{fig:BBU_IU_Scenrio}  illustrates such a scenario, where devices can be utilized for near real-time traffic monitoring via a digital twin, and real-time actuation of traffic signals}. Due to its intermittent behaviour, we consider that the monitoring device transmits its updates to the cellular network through a grant-free access mechanism. For ease of exposition, we will henceforth refer to this device as the \emph{intermittent user}. \textcolor{black}{The \gls{bs} allocates available wireless resources among the intermittent user and a broadband user that generates data continuously\cite{Ley23Asilomar,OJCOMMS@fedrico}.} Moreover, a frame-based communication model is considered, where uplink and downlink transmissions are multiplexed in time. For the considered model, we analyse the performance metrics of the intermittent user and compare them with those obtained from the model employed in conventional studies on \textcolor{black}{real-time remote tracking}. {To this end, we consider \gls{tare}, \gls{tacae} and \gls{udc} as performance metrics for the intermittent user. Here, we define \gls{udc} as the average number of re-transmissions needed to successfully deliver an update. We show that the desirably low \gls{tare} and \gls{tacae} performance of the communication model in conventional studies is achieved at the expense of a disproportionately high \gls{udc}.} We then study performance trade-offs between the intermittent user and the broadband user, whose metrics are throughput and energy efficiency, by sharing the available wireless resources using \gls{fdma} and \gls{noma} schemes. \textcolor{black}{We show that \gls{noma} achieves a better performance trade-off between the two users, particularly when energy efficiency is the performance measure for the broadband user.}

\section{System model}\label{Sys_Mod}
We consider an uplink scenario wherein a broadband user and an intermittent user are allocated wireless resources to communicate with the \gls{bs}. The users and the \gls{bs} are equipped with a single antenna. The users are indexed by $m\in\{1,2\}$, with the broadband user indexed as $m=1$ and the intermittent user as $m=2$. The available wireless resources are within a frequency band of bandwidth $B$ Hz\cite{Ley23Asilomar}. This bandwidth $B$ is divided into three sub-bands: $1$) $B_{1}$ reserved for the broadband user, $2$) $B_{2}$ reserved for the intermittent user, and $3$) $B_{3}$ to be shared by both users, such that $B_{1}+B_{2}+B_{3}=B$. Let $\alpha_{m,i}\in\{0,1\}$ indicate the allocation of user $m$ to sub-band $i\in\{1,2,3\}$, such that $\alpha_{m,i}=1$ if user $m$ is assigned to sub-band $i$, and $\alpha_{m,i}=0$ otherwise. We consider a time-slotted communication system with $T_{s}$ as the duration of each slot. The communication is segmented into frames, with each frame consisting of $T_{F}$ consecutive time slots. Subsequently, we detail the multiple access schemes as follows.\footnote{\textcolor{black}{We exclude \gls{tdma} because it either restricts the choice of the sampling policy or renders the analysis intractable, requiring joint design of both the sampling policy and resource allocation.}}
\begin{enumerate}
    \item \textit{\Gls{fdma}:} Users are allocated non-overlapping frequency sub-bands, with $B_{1}$ containing slots reserved for the broadband user and $B_{2}$ for the intermittent user. Hence, signal overlap between the users is avoided by  setting $\alpha_{1,1}=\alpha_{2,2}=1$, $B=B_{1}+B_{2}$, and $B_{3}=0$\cite{Ley23Asilomar}. 
    \item \textit{\Gls{noma}:} Both users are allocated the entire bandwidth by setting $\alpha_{1,3}=\alpha_{2,3}=1$, $B_{1}+B_{2}=0$ and $B_{3}=B$. This leads to a complete signal overlap between the users when both transmit in the same time slot\cite{Ley23Asilomar}. 
\end{enumerate}
\vspace{-0.1 cm}
\subsection{Transmission Model}\label{SysMod_TransMod}
\begin{figure}
    \centering
    \begin{tikzpicture}
\pgfmathsetmacro\y{1.6}
\pgfmathsetmacro\w{0.4}
\pgfmathsetmacro\h{0.8}
\pgfmathsetmacro\b{0.4}
\node[anchor=south] at (5*\w,1.02*\h){\textbf{FDMA}};
\foreach \x in {0,1,...,8}{
    \filldraw[Greys-H, fill=Set3-E] (\x*\w,\h*\b) rectangle ({\w*(\x+1)},\h);
    \filldraw[Greys-H, fill=Set3-B] (\x*\w,0) rectangle ({\w*(\x+1)},\h*\b);
    }
\draw[|-|, xshift=-20pt, Greys-H] (0,0)--(0,\h)node[midway, left, font=\scriptsize, align=center]{Bandwidth\\ $B$};
\draw[|-, xshift=-4pt, Greys-H] (0,0)--(0,\h*\b)node[midway, left, font=\scriptsize]{$B_2$};
\draw[|-|, xshift=-4pt, Greys-H] (0,\h*\b)--(0,\h)node[midway, left, font=\scriptsize]{$B_1$};

\filldraw[Greys-H, fill=Greys-E] (9*\w,0) rectangle ({\w*(10)},\h);
\node[Greys-J, font=\scriptsize] at ({\w*9.5},\h*0.5){FB};
\node[Greys-J, font=\scriptsize] at ({\w*10.8},\h*0.5){$\cdots$};
\draw[|-|, yshift=-4pt, Greys-H] (0,0)--(\w,0)node[midway, below, font=\scriptsize]{Time slot};

%%%%%%%%%%  TDMA
\iffalse
\begin{scope}[yshift = -1cm*\y]
\node[anchor=south, align=center] at (5*\w,+1.02*\h){\textbf{TDMA}};
\foreach \x in {0,1,3,4,6,7}{
    \filldraw[Greys-H, fill=Set3-E] (\x*\w,0) rectangle ({\w*(\x+1)},\h);
}
\foreach \x in {2,5,8}{
    \filldraw[Greys-H, fill=Set3-B] (\x*\w,0) rectangle ({\w*(\x+1)},\h);
    }
\draw[|-|, xshift=-4pt, Greys-H] (0,0)--(0,\h)node[midway, left, font=\scriptsize]{$B_i$}; 

\draw[|-|, xshift=-20pt, Greys-H] (0,0)--(0,\h)node[midway, left, font=\scriptsize, align=center]{Bandwidth\\ $B$};
\filldraw[Greys-H, fill=Greys-E] (9*\w,0) rectangle ({\w*(10)},\h);
\node[Greys-J, font=\scriptsize] at ({\w*9.5},\h*0.5){FB};
\node[Greys-J, font=\scriptsize] at ({\w*10.8},\h*0.5){$\cdots$};

\end{scope}
\fi

%%%%%%%%%%  NOMA
\begin{scope}[yshift = -1cm*\y]
\node[anchor=south, align=center] at (5*\w,+1.02*\h){\textbf{NOMA}};
\foreach \x in {0,1,...,8}{
    \filldraw[Greys-H, fill=Pastel2-A] (\x*\w,0) rectangle ({\w*(\x+1)},\h);
}
\draw[|-|, xshift=-4pt, Greys-H] (0,0)--(0,\h)node[midway, left, font=\scriptsize]{$B_3$}; 

\draw[|-|, xshift=-20pt, Greys-H] (0,0)--(0,\h)node[midway, left, font=\scriptsize, align=center]{Bandwidth\\ $B$};
\draw[|-|, yshift=-4pt, Greys-H] (0,0)--(10*\w,0)node[midway, below, font=\scriptsize]{Frame};
\filldraw[Greys-H, fill=Greys-E] (9*\w,0) rectangle ({\w*(10)},\h);
\node[Greys-J, font=\scriptsize] at ({\w*9.5},\h*0.5){FB};
\node[Greys-J, font=\scriptsize] at ({\w*10.8},\h*0.5){$\cdots$};

\end{scope}

%%%%%%% Legend
\begin{scope}[xshift=4.8cm, yshift=0.5cm, font=\scriptsize]
\pgfmathsetmacro\y{0.5}
\pgfmathsetmacro\w{0.2}
\pgfmathsetmacro\h{0.2}
\draw[thick] (0.1,-2*\y-\w) rectangle (2,\h+\w);
\filldraw[Greys-G, fill=Set3-E] (\w,0) rectangle (2*\w,\h);
\node[xshift=2pt,anchor=west] at (2*\w,0.5*\h){Broadband};
\filldraw[Greys-G, fill=Set3-B] (\w,-\y) rectangle (2*\w,\h-\y);
\node[xshift=2pt,anchor=west] at (2*\w,0.5*\h-\y){Intermittent};
\filldraw[Greys-G, fill=Pastel2-A] (\w,-2*\y) rectangle (2*\w,\h-2*\y);
\node[xshift=2pt,anchor=west] at (2*\w,0.5*\h-2*\y){Shared};
\end{scope}
\end{tikzpicture}\vspace{-2em}
    \caption{Considered frame structure with \gls{fdma} and \gls{noma}.}
    \label{fig:frame_access_mechanisms}\vspace{-1.6em}
\end{figure}
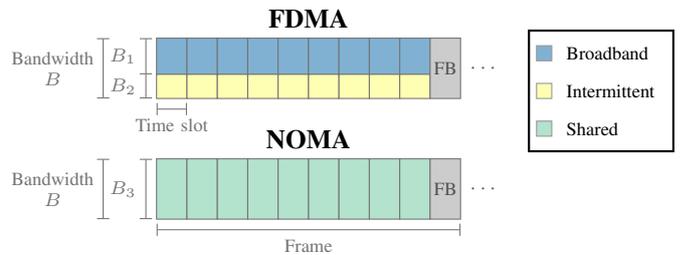

The first $T_{F}-1$ slots are reserved for uplink transmission, while the last slot is reserved for downlink transmission to provide feedback, \gls{ack} or \gls{nack}, to the users from the \gls{bs}. Here, the time slots and frames are indexed by $t\in\{0,1,\ldots \}$ and $f\in\{0,1,\ldots \}$, respectively. Fig.~\ref{fig:frame_access_mechanisms} illustrates the frame structure with \gls{fdma} and \gls{noma} schemes\cite{Ley23Asilomar}. \textcolor{black}{Next, we outline the transmission policy of the broadband user and the intermittent user, respectively.} 
\subsubsection{Broadband user}\label{SysModTransMod_BBU}The user segments its transmit data into packets and employs a forward error correction (FEC) mechanism using an ideal rate-less packet-level coding, wherein blocks of $K$ source packets are encoded into blocks of linearly independent packets. \textcolor{black}{An encoded block can span over multiple frames. Subsequently, the \gls{bs} is able to decode a block of source packets once it has successfully received $K$ encoded packets representing the source block. Successful/failed decoding of a source block is informed to the user by the \gls{bs} using feedback at the end of a frame. If the user receives an \gls{ack} through feedback, then the user proceeds to transmit encoded packets of the next source block, otherwise, the user keeps transmitting the encoded packets of the current source block. Such transmission strategy ensures that the broadband user is guaranteed to achieve reliability $1$.}
\subsubsection{Intermittent user}\label{SysModTransMod_IU} 
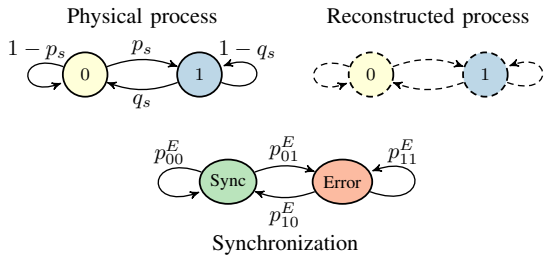
\begin{figure}[t]
    \centering
    \resizebox{0.4\textwidth}{!}{\begin{tikzpicture}[->, >=stealth', auto, semithick, state/.append style={minimum size=20pt, font=\footnotesize, inner sep=1pt, ellipse}]
\pgfmathsetmacro{\xshft}{1.8}

 \node[state, thick,draw]    (1)[fill=YlGnBu-A] at (0,0)  {$0$};
\node[state, thick,draw]    (2)[ fill=Blues-E] at (\xshft,0)  {$1$};
\node at(0.5*\xshft,0.9){Physical process};

\path
 (1) edge[bend left=20]node[pos=0.5,above, inner sep=2pt]{$p_s$}	              (2)
 (2) edge[bend left=20]node[pos=0.5,below, inner sep=2pt]{$q_s$}	              (1)
  (1) edge[in=200,out=160,loop]node[pos=0.25,above, inner sep=2pt]{$1-p_s$}	             (1)
 (2) edge[in=20,out=340,loop]node[pos=0.75,above,inner sep=2pt]{$1-q_s$}	             (2);

 \begin{scope}[xshift=2.5*\xshft*1cm, Greys-M, densely dashed]
 \node[state, thick, draw]    (dt1)[fill=YlGnBu-A] at (0,0)  {$0$};
\node[state, thick, draw]    (dt2)[ fill=Blues-E] at (\xshft,0)  {$1$};
\node at(0.5*\xshft,0.9){Reconstructed process};

\path[]
 (dt1) edge[bend left=20]              (dt2)
 (dt2) edge[bend left=20]              (dt1)
  (dt1) edge[in=200,out=160,loop]	             (dt1)
 (dt2) edge[in=20,out=340,loop]	             (dt2);
 \end{scope}

 \begin{scope}[xshift=1.25cm*\xshft, yshift=-1.7cm]
 \node[state, thick, draw]    (s1)[fill=GnBu-E] at (0,0)  {Sync};
\node[state, thick, draw]    (s2)[ fill=Reds-D] at (\xshft,0)  {Error};
\node at(0.5*\xshft,-1){Synchronization};

\path[]
 (s1) edge[bend left=20]node[pos=0.5,above,inner sep=2pt]{$p^E_{01}$}              (s2)
 (s2) edge[bend left=20]node[pos=0.5,below,inner sep=2pt]{$p^E_{10}$}              (s1)
  (s1) edge[in=200,out=160,loop]node[pos=0.25,above,inner sep=2pt]{$p^E_{00}$}	             (s1)
 (s2) edge[in=20,out=340,loop]node[pos=0.75,above,inner sep=2pt]{$p^E_{11}$}	             (s2);
 \end{scope}

\end{tikzpicture}}
\vspace{-8pt}\vspace{-0.1 cm}
    \caption{Two-state \gls{dtmc} of the physical process being monitored, its reconstruction at the BS, and the state of the system vis-à-vis synchronization.}
    \label{fig:dtmc}\vspace{-0.5 cm}
\end{figure}

We model the information source at the user by an ergodic two-state \gls{dtmc} $\{X_{t},\,t\in\mathbb{Z}_{0}^{+}\}$, where the state $X_{t}$ of the source at time slot $t$ can be $0$ or $1$~\cite{pappas@dtmcTWC}. \textcolor{black}{In this model, the states represent distinct sets of values collected from one or multiple sensors, aggregated at a single communication device which, for ease of exposition, are simply denoted as $0$ or $1$ in the two-state \gls{dtmc}\cite{pappas@dtmcTWC}}. The self-transition probabilities for states $0$ and $1$ are denoted as $1-p_{s}$ and $1-q_{s}$, respectively. As a result,  $\mathbb{P}\left(X_{t+1}=X_{t}\right) = \mathbbm{1}\left(X_{t}=0\right)\left(1-p_{s}\right)+\mathbbm{1}\left(X_{t}=1\right)\left(1-q_{s}\right)$, with $\mathbbm{1}(\cdot)$ denoting the indicator function. The user observes the process $X_{t}$ and informs the \gls{bs} about its state by sending updates over a wireless channel. To that end, the user will generate an update $X_{t}$ through sampling, and transmit the update to the \gls{bs} in a data packet of length $L$. {Note that the structure of the frame, illustrated in Fig.~\ref{fig:frame_access_mechanisms}, allows for the instantaneous transmission of an update only in the first $T_{F}-1$ slots of the frame. Moreover, we assume that the intermittent user has a transmission queue of length 1, so any new update replaces the previous packet in the queue.} The actions of sampling and transmission at time slot $t$ are denoted by $\alpha_{t}^{\textrm{s}}\in \{0,1\}$ and $\alpha_{t}^{\textrm{tx}}\in \{0,1\}$, respectively, where $\alpha_{t}^{a}=1$, $\,a\in \{\textrm{s},\textrm{tx}\}$, indicates that the action is taken at time slot $t$, and $\alpha_{t}^{a}=0$ otherwise. Next, if the \gls{bs} successfully receives the packet, it updates the state of the reconstructed source instantaneously. {The feedback from the \gls{bs} to the intermittent user, and subsequently the re-transmission of the failed update, depends on the sampling policy employed for process monitoring (see Section~\ref{ResAll_Samp}). Notwithstanding, if the sampling policy requires feedback and multiple updates are sent within a frame, feedback will be provided only for the latest update. Fig.~\ref{fig:dtmc} illustrates the process monitoring of the considered source process. As highlighted in Section \ref{Intro_Section}, the reconstructed process at the \gls{bs} can be utilized to construct a digital twin or effectuate real-time actuation.} 
\subsection{Physical Layer Model}\label{SysMod_PHYLayer}
We denote the channel coefficient between the \gls{bs} and user $m$ at time slot $t$ as $h_{m,t}\in \mathbb{C}$. The channel coefficient is modeled as a random variable that accounts for large-scale and small-scale fading losses. The small-scale fading loss is modeled as a \gls{cscg} random variable with zero mean and unit variance. On the other hand, the large-scale fading loss is a function of the speed of light $c$, the carrier frequency $f_{c}$, the distance between the \gls{bs} and user $m$, $d_{m}$, the path loss exponent $\eta$, and the antenna gains at the \gls{bs} and the user, $G_{b}$ and $G_{m}$, respectively. Thus, the large-scale fading loss, $\beta_{m}$, can be computed as 
\begin{equation}\label{eq:LSF}
    \beta_{m}=\mathbb{E}\{\lvert h_{m,t}\rvert^{2}\}=\frac{G_{m}G_{b}c^{2}}{\left(4\pi f_{c}\right)^{2}d_{m}^{\eta}},\:m\in\{1,2\}.
\end{equation}

By $x_{m,t}\in\mathbb{C}$ we denote as the transmit signal between user $m$ and the \gls{bs} at time slot $t$, and by $P_{m,t}\in [0,P_{\textrm{max}}]$ as the transmission power of user $m$. Subsequently, the received signal at the \gls{bs} during uplink transmission in the $i$-th sub-band  and $t$-th time slot can be written as 
\begin{equation}\label{eq:rx_sig}
    y_{i,t}= h_{1,t}x_{1,t}\alpha_{1,i} +h_{2,t}x_{2,t}\alpha_{2,i} + w_{i,t},\; \forall i\in\{1,2,3\},
\end{equation}
where $w_{i,t}$ is the \gls{cscg} distributed additive noise with zero mean and variance $\sigma_{i}^{2}=B_{i}\kappa T_{w}10^{N_{f}/10}$. Here, $\kappa$ is the Boltzmann constant, $T_{w}$ is the system noise temperature, and $N_{f}$ is the noise figure in dB. Hence, the \gls{sinr} for the signal of user $m$ in the $i$-th sub-band and $t$-th time slot can be computed as
\begin{equation}\label{eq:sinr_mit}
    \gamma_{m,i,t}=\frac{\lvert h_{m,t}\rvert^{2}P_{m,t}\alpha_{m,i}}{\lvert h_{n,t}\rvert^{2}P_{n,t}\alpha_{n,i}+\sigma_{i}^{2}},\,\, m,n\in\{1,2\}\,\wedge\,m\neq n.
\end{equation}
Assuming the interference can be treated as Gaussian noise, we define a threshold for decoding the signal of user $m$ at time slot $t$ in sub-band $i$ as the function of the rate $r_{m}$, as
\begin{equation}\label{eq:sinr_min}
\gamma_{m,i}^{\textrm{min}}\left(r_{m}\right)=2^{r_{m}/B_{i}}-1,
\end{equation}
such that if $\gamma_{m,i,t}\geq \gamma_{m,i}^{\textrm{min}}\left(r_{m}\right)$, then the packet of user $m$ is successfully decoded at time slot $t$; we denote the corresponding probability of successful decoding by
\begin{equation}\label{eq:success_prob}
p_{m,i,t}=\mathbb{P}\left(\gamma_{m,i,t}\geq \gamma_{m,i}^{\textrm{min}}\left(r_{m}\right)\right).
\end{equation}
Further, we introduce the error probability for user $m$ as $\epsilon_{m,i}$, defining it as the probability of a transmission error in the absence of interference from the other user $n\neq m$, given by 
\begin{IEEEeqnarray}{C}\label{eq:erasure_prob}
    \epsilon_{m,i}= \mathbb{P}\left(\gamma_{m,i,t}\leq \gamma_{m,i}^{\textrm{min}}\left(r_{m}\right)\!:\alpha_{m,i}=1\wedge P_{n,t}\alpha_{n,i}=0\right)\!.\IEEEeqnarraynumspace
\end{IEEEeqnarray}
{The \gls{bs} aims to decode the received signals instantaneously, i.e., in the same time slot it received the packet.} While decoding the signal of user $m$ in the \gls{fdma} regime is straightforward, we consider a \gls{sic} decoder in conjunction with the \emph{capture effect} for \gls{noma}\cite{OJCOMMS@fedrico,Ley23Asilomar}. The success or failure of the decoding process is dictated by the \gls{sinr} defined in \eqref{eq:sinr_mit}. \textcolor{black}{The \gls{bs} attempts to decode the received signal and, if successful, applies SIC to decode the remaining signal (if any), ensuring both users can be decoded if the \gls{sinr} and \gls{snr} thresholds are met.}
\section{Performance Metrics}\label{Perf_Metrics}
In this section, we derive the expression for metrics relevant to the performance of the broadband and intermittent user.
\subsection{Broadband User}\label{PerfMetrics_BBU}
For the broadband user, we consider throughput and energy efficiency to be the performance metrics. To this end, the transmission rate $r_{1}$ is selected as the minimum of: 1) the maximum rate $r_{1}^{\textrm{max}}$ and 2) the maximum achievable data rate for the target error probability $\epsilon_{1}^{*}$, expressed as
\begin{equation}\label{eq:BBuser_r1}
    r_{1}=\max\{r\in (0,r_{1}^{\textrm{max}}]:\epsilon_{1,i}\left(r\right)=\epsilon_{1}^{*}, P_{1,t}\leq P_{m}^{\textrm{max}}\}.
\end{equation}
For the desired rate $r_{1}$ and the target error probability $\epsilon_{1}^{*}$, $P_{1,t}$ can be calculated by assuming absence of interference in \eqref{eq:sinr_mit} and then utilizing equations \eqref{eq:erasure_prob} and \eqref{eq:success_prob}, given by 
\begin{equation}\label{eq:Power_BBUser}
P_{1,t}=\min\left(\frac{\left(2^{r_{1}/B_{i}}-1\right)\sigma_{i}^{2}}{\mathbb{E}[|h_{1}|^{2}]\log(\epsilon_{1}^{*}-1)},P_{\textrm{max}}\right),
\end{equation}
where it is assumed that the broadband user has statistical knowledge of its channel, i.e., of $\mathbb{E}[|h_{1}|^{2}]$.
Denoting the random variable of the number of frames needed to decode the block of $K$ source packets transmitted by the user as $F(K)$, the throughput of the broadband user is calculated as
\begin{equation}\label{eq:BBuser_Throughput}
    S_{B}=r_{i}K\left(T_{F}-1\right)/\left(\mathbb{E}\{F(K)\}T_{F}\right),
\end{equation}
and the energy efficiency is calculated as $S_{B}/P_{1,t}$.
\subsection{Intermittent User}\label{PerfMetrics_IU} 
We consider context-dependent and cost-aware performance metrics for the intermittent user \cite{pappas@dtmcTWC,pappas@dtmc}. These rely on the \gls{bs} constructing an estimate of the original source state $X_{t}$ at time slot $t$, denoted by $\widehat{X}_{t}$.
{We consider that $\widehat{X}_{t}=X_t$ whenever an update is received at time slot $t$ and $\widehat{X}_{t}=\widehat{X}_{t-1}$ otherwise.} These metrics are described in the following.
\subsubsection{\gls{tare}}\label{PerfMetricsIU_RCE}   The real-time reconstruction error is %defined as the difference between the original source state $X_{t}$ and the reconstructed source state $\widehat{X}_{t}$ at time slot , 
calculated as  $E_{t}=\mathbbm{1}\left(X_{t}\neq\widehat{X}_{t}\right)\in\{0,1\}$\cite{pappas@dtmcTWC}. As shown in Fig.~\ref{fig:dtmc}, $E_{t}$ can be construed as an indicator for the state of the system, with $E_{t}=1$ being the erroneous state and $E_{t}=0$ being the synced state.  Subsequently, for an observation interval of $[1,T_{o}]$, with $T_{o}$ being a large number, the \gls{tare} is defined as 
\begin{equation}\label{eq:Iuser_ARE}
    \bar{E}=\lim_{T_{o}\rightarrow \infty} \frac{1}{T_{o}}\sum_{t=1}^{T_{o}}{E}_{t}= \frac{1}{T_{o}}\sum_{t=1}^{T_{o}}\mathbbm{1}\left(X_{t}\neq\widehat{X}_{t}\right).
\end{equation}
Interestingly, the evolution of $E_{t}$ can be described by a Markov chain as well, as shown in Fig.~\ref{fig:dtmc}; with transition probabilities defined as $p_{ji}^{E}=\mathbb{P}\left(E_{t+1}=j\mid E_{t}=i\right),\, \forall i,j\in\{0,1\}$. For ease of understanding, we derive the general expression of the transition probabilities $p_{00}^{E}$ and $p_{11}^{E}$ as follows. 
\begin{equation}\label{eq:Iuser_REP00}
    p_{00}^{E}=\sum_{i=0}^{1} \mathbb{P}\left(E_{t+1}=0\mid E_{t}=0,\,X_{t}=i\right)\mathbb{P}\left(X_{t}=i\right),
\end{equation}
with $\mathbb{P}\left(X_{t}=i\right)=\mathbbm{1}(i=0)\frac{q_{s}}{p_{s}+q_{s}}+\mathbbm{1}(i=1)\frac{p_{s}}{p_{s}+q_{s}},\,\forall i\in\{0,1\}$, calculated using stationary distribution of $X_{t}$. To compute \eqref{eq:Iuser_REP00}, we first define an indicator event for successful decoding of a transmitted packet of the intermittent user as $\alpha^{\textrm{d}}$ such that $p_{2,i,t}=\mathbb{P}\left(\alpha_{t}^{\textrm{d}}=1\right)$. Naturally, $\mathbb{P}\left(\alpha_{t}^{\textrm{d}}=0\right)=1-p_{2,i,t}$. Next, with $j\in \{0,1\}$, we denote probabilities of joint sampling, transmission, and decoding events at time $t+1$ as $p_{\textrm{stx}}^{j}=\mathbb{P}\left(\alpha_{t+1}^{\textrm{s}}=1,\alpha_{t+1}^{\textrm{tx}}=j\right)$, and $p_{\textrm{stx}}^{\textrm{d}_{j}}=\mathbb{P}\left(\alpha_{t+1}^{\textrm{s}}=1, \alpha_{t+1}^{\textrm{tx}}=1, \alpha_{t+1}^{\textrm{d}}=j\right)$; Section \ref{SysModTransMod_IU} delineates events $a_{t}^{\textrm{s}}$ and $a_{t}^{\textrm{tx}}$ in detail. Subsequently, we obtain
\begin{equation}\label{eq:Iuser_REP000}
\mathbb{P}\left(E_{t+1}=0\mid E_{t}=0,\,X_{t}=0\right)=\left(1-p_{s}\right)+p_{s}p_{\textrm{stx}}^{\textrm{d}_{1}}.
\end{equation}
Similarly, $\mathbb{P}\left(E_{t+1}=0\mid E_{t}=0,\,X_{t}=1\right)$ can be calculated. In regard to $p_{11}^{E}$, we have
\begin{equation}\label{eq:Iuser_REP110}
\mathbb{P}\left(E_{t+1}=1\mid E_{t}=1,X_{t}=0\right)
=(1-p_{s})(p_{\textrm{stx}}^{0}+p_{\textrm{stx}}^{\textrm{d}_{0}}).   
\end{equation}
To obtain $\mathbb{P}\left(E_{t+1}=1\mid E_{t}=1,\,X_{t}=1\right)$, we simply replace $p_{s}$ in \eqref{eq:Iuser_REP110} by $q_{s}$. Evidently, the transition probabilities depend on the sampling policy as well as the multiple access scheme. 
\subsubsection{\gls{tacae} \cite{pappas@dtmcTWC}}\label{PerfMetricsIU_CAE} Building on the real-time reconstruction error metric, we consider the cost of actuation error to capture the significance of the reconstruction error at the point of actuation. {To this end, we denote $C_{i,j}$ as the cost of being in state $j$ at the reconstructed source ($\widehat{X}_t=j$) when the original source is in state $i$ (${X}_t=i$), with $C_{i,i}=0$.} Moreover, we assume $C_{i,j}\neq C_{j,i}$ as different errors have different repercussions on the actuation. Finally, we assume that $C_{i,j}$, $\forall i,j$, is static over the entire observation interval. Following the above discussion, we calculate the \gls{tacae} as
\begin{equation}\label{eq:Iuser_AE}
    \bar{C}_{A}= \pi_{\left(0,1\right)}C_{0,1} + \pi_{\left(1,0\right)}C_{1,0},
\end{equation}
where $\pi_{\left(i,j\right)},\,\forall i,j\in\{0,1\}$ is obtained from the stationary distribution of the \gls{dtmc}, describing the joint status of the source state $X_{t}$ and the reconstruction error state $E_{t}$ at $t$.
\section{Sampling Policy and Multiple Access Schemes}\label{ResAll_Samp}
In this section we delineate the sampling policy for process monitoring at the intermittent user and highlight its effect on the performance metrics. {It should be emphasized that the choice of the sampling policy is closely intertwined with the choice of the multiple access scheme. Keeping this in mind, we consider the semantics-aware sampling policy which triggers sample generation in two cases\cite{pappas@dtmcTWC}. First, whenever a change in the state of the source between two consecutive slots is observed, i.e., ${X}_{t+1}\neq{X}_{t}$. Second, when the system is known to be in erroneous state, i.e., $\widehat{X}_{t}\neq {X}_{t}$\cite{pappas@dtmcTWC}. Such a policy integrates well with \gls{fdma} and \gls{noma}. Moreover, unlike uniform and change-aware sampling policies, the semantics-aware policy requires feedback from the receiver to the transmitter to trigger sample generation\cite{pappas@dtmcTWC}. Additionally, feedback is used to re-transmit failed update deliveries. The sampling policy, together with the frame structure, dictates the probability of sampling and transmission events and in turn the value of $p_{\textrm{stx}}^{j}$ in equations \eqref{eq:Iuser_REP000} and \eqref{eq:Iuser_REP110}. On the other hand, together with the choice of multiple access scheme, which dictates the probability of successful decoding $p_{2,i,t}$, it determines the value of  $p_{\textrm{stx}}^{\textrm{d}_{j}}$. To this end, we delineate the calculation of $p_{2,i,t}$ for \gls{fdma} and \gls{noma} schemes. We assume the intermittent user has no knowledge of its channel due to its sporadic access behavior and, therefore, always transmits at power $P_{\textrm{max}}$.} 
\subsubsection{FDMA} As the users are allocated orthogonal sub-bands, the signal of the intermittent user is not subject to interference from the broadband user. Therefore, $p_{2,i,t}$ for the transmitted packet at time slot $t$ can be calculated from \eqref{eq:success_prob}, where $\gamma_{2,i,t}$, obtained using \eqref{eq:sinr_mit}, will be devoid of the interference term. 
\subsubsection{NOMA}Since users share time-frequency resources in \gls{noma} and the  capture effect can occur, the successful decoding probabilities of both users' signals at time slot $t$ become interdependent. Therefore, to calculate $p_{i,2,t}$ for the intermittent user, we first define two possible outcomes
\begin{itemize}
    \item $\mathcal{I}$: The signal of the intended user gets successfully decoded. Either $1$) the signal of the intended user has a sufficiently high \gls{sinr} to be decoded directly, or $2$) the signal from the other user is decoded first, its interference is eliminated using \gls{sic}, and then the intended signal has a sufficiently high \gls{snr} and gets decoded.
    \item $\mathcal{E}$: The signal of the intended user has insufficient \gls{snr} to be decoded.
\end{itemize}
{Evidently, the transmission from the intermittent user will get successfully decoded in two cases: $\left(\mathcal{I},\mathcal{I}\right)$ and $\left(\mathcal{E},\mathcal{I}\right)$, where ordered pairs $(\cdot,\cdot)$ describe possible outcomes for the signals of both users when they overlap.} Subsequently, the probability of successful decoding of the signal of the intermittent user can be calculated as $p_{2,i,t}=\pi_{\mathcal{I}\mathcal{I}}+\pi_{\mathcal{E}\mathcal{I}}$, where both $\pi_{\mathcal{I}\mathcal{I}}$ and $\pi_{\mathcal{E}\mathcal{I}}$ can be obtained from Appendix A in \cite{OJCOMMS@fedrico}.
\section{Results}
In this section, we evaluate the performance of the broadband user and the intermittent user sharing resources in the considered frame-based communication model. Here, the broadband user is located at a distance of $50\,$m, whereas the intermittent user is located at a distance $d\,$ from the \gls{bs}. The system parameters are listed in Table \ref{tab:Sim_table}.  
\begin{table}
	\caption{Simulation parameters}\vspace{-0.3cm}
	    \label{tab:Sim_table}\centering
	\begin{tabular}{l l l}
		\toprule[0.4mm]
		\textbf{Parameter} & \textbf{Symbol}& \textbf{Value}\\
  		\toprule[0.4mm]
   Broadband user erasure probability & $\epsilon^{*}$& $0.1$\\
   Broadband user maximum data rate & $r_{1}^{\textrm{max}}$& $5$ Mbps\\
   Maximum transmission power & $P_{\textrm{max}}$ & $200$ mW \\
   Antenna gains & $G_{t}G_{r}$& $10$\\
   Time slot duration & $T_{s}$ & $1$ ms\\
   Carrier frequency &$f_{c}$& $2$ GHz\\
   System bandwidth & $B$ & $1$ MHz\\
   Noise temperature & $T_{w}$& $190$ K\\
   Noise figure & $N_{f}$ & $5$ dB\\
   Frame length & $T_{F}$ & $10$\\
   Broadband user source block length & $K$ & $32$\\
   Intermittent user packet length & $L$ & $128\,\textrm{B}$\\
   \gls{bs}-intermittent user distance  & $d$ & $\{100, 200, 400\}\,$m\\
   Path loss exponent & $\eta$ &$2.6$\\
   Cost of actuation error coefficients & $C_{i,j}$ &$C_{0,1}=5,\,C_{1,0}=1$\\
	\bottomrule[0.4mm]	\end{tabular}\vspace{-0.4cm}
\end{table}
To begin with, we analyse the performance of the intermittent user for \gls{fdma} frames (i.e., $B_{3}=0$), with increasing value of $B_{2}$. {For comparison, we consider the frame-less communication model assumed in conventional studies as the baseline~\cite{pappas@dtmcTWC}. Such model `idealistically' assumes instantaneous transmission of the updates and feedback, hence, is denoted as the `Idealistic' model throughout this section. The time-averaged reconstruction error, \gls{tare}, for the Idealistic model with the semantics-aware sampling policy can be analytically calculated as}
\begin{equation}\label{eq:Iuser_RE_Theo}
 \bar{E}=\frac{2\,p_{s}q_{s}\left(1-p_{2,i,t}\right)}{p_{2,i,t}\left(p_{s}+q_{s}\right)+4p_{s}q_{s}\left(1-p_{2,i,t}\right)}.
\end{equation}
However, deriving the analytical expression of the \gls{tare} for the model in Section \ref{SysMod_TransMod}, denoted as the `Frame-Based' model, is only tractable for sampling policies that do not require feedback, e.g., uniform and change-aware. This is due to the constraints of the frame structure on the transmission of feedback, discussed in Section \ref{Sys_Mod}. Since the considered semantics-aware sampling policy requires feedback, deriving an analytical expression similar to \eqref{eq:Iuser_RE_Theo} for the Frame-Based model becomes intractable. Therefore, we numerically evaluate the performance of the intermittent user in the Frame-Based model by simulating more than $100000$ frames.
\begin{figure}[!t]
\centering 
\includegraphics[width=0.96\columnwidth]{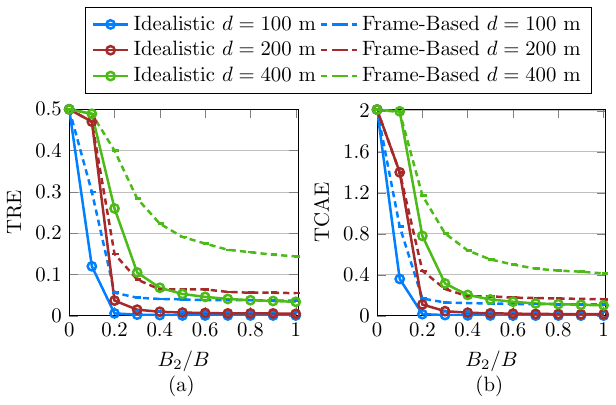}\vspace{-0.35cm}\\
\caption{\gls{tare} and \gls{tacae} of the intermittent user}\vspace{-0.50cm}
\label{fig:ReconstructionError_IU}
\end{figure}
\par Fig.~\ref{fig:ReconstructionError_IU}$\left(\textrm{a}\right)$ illustrates the \gls{tare} with increasing $B_{2}=\{0.1B,0.2B,\ldots,B\}$; for source transition probabilities $\left(p_s=0.1,\,q_s=0.15\right)$ and distance $d=\{100, 200, 400\}\,$m. The relatively small values of $p_s$ and $q_s$ imply that the \gls{dtmc} source is changing slowly, whereas varying values of $d$ are considered to observe the \gls{tare} for varying decoding probabilities. The figure illustrates that the successful decoding probability increases with increasing $B_{2}$ and, consequently, the \gls{tare} at the \gls{bs} decreases. Moreover, the farther the intermittent user is from the \gls{bs}, the larger $B_{2}$ needs to be to reduce the \gls{tare}. As expected, with increasing $B_{2}$ the decline in \gls{tare} is steeper for the Idealistic model. This gets more pronounced with increasing $d$. However, the performance gain of the Idealistic model over the Frame-Based model comes at the expense of very  high \gls{udc}, as illustrated in Table \ref{tab:CAE_RE_SC} for parameters $B_{2}=0.4\,B$ and $d=400$. The table shows that while \gls{tare} for the Idealistic model is $70\%$ less than that in Frame-Based model, the \gls {udc} is $134\%$ more. Similar behaviour is observed with the \gls{tacae} metric; albeit the difference in performance is narrower, as observed in Fig.~\ref{fig:ReconstructionError_IU}$\left(\textrm{b}\right)$.  
\begin{table}[!b]
\vspace{-0.5cm}\centering
\caption{Intermittent user Performance metrics, $B_{2}=0.4\,B$ and $d=400$}\vspace{-0.0cm}
   {\begin{tabular}{|l|c|c|}
 \hline
 \multirow{3}{*}{} & \multicolumn{1}{c|}{Idealistic} & 
    \multicolumn{1}{c|}{Frame-Based}\\
   \hline  
  {\gls{tare}} &  {$0.068$} &  {$0.223$}\\ 
 \hline
{\gls{tacae}} &  {$0.204$} &  {$0.638$}\\
 \hline
 {\gls{udc}} &  {$0.614$} &  {$0.263$}\\
 \hline
\end{tabular}}
    \label{tab:CAE_RE_SC}
\end{table}
\par The impact of the \gls{udc} becomes further pronounced in the case of a fast changing \gls{dtmc} process, as illustrated in Table~\ref{tab:Slow_and_fast_DTMC} that compares the performance metrics of the intermittent user for a slow process $\left(p_s=0.1,\,q_s=0.15\right)$ and a fast process $\left(p_s=0.2,\,q_s=0.7\right)$, for $B_{2}=0.4B$ and $d=400$. It can be observed that as the \gls{dtmc} process becomes fast changing, both error metrics increase for the Idealistic model. In comparison, for the Frame-Based model, the \gls{tare} remains the same and the \gls{tacae} decreases.
Such discrepancy is by virtue of the fact that delayed feedback is the limiting factor for the Frame-Based model. Therefore, as the process becomes fast changing, it leads to more frequent updates which in turn mitigates the inimical impact of delayed feedback. {As a result, while the \gls{udc} decreases by $16\%$ and $54\%$ for the Idealistic and Frame-Based models, respectively, due to frequent updates, performance metrics degrade for the former model but improve for the latter. Alternatively, the results also suggest that the desired error-metric performance for the Frame-Based model can be achieved by adjusting the frame length, with only a modest change in the \gls{udc}.}
\begin{table}
\centering
\caption{Performance of the intermittent user with source variability}\vspace{-0.15cm}
   \resizebox{\columnwidth}{!}{\begin{tabular}{|l|c|c|c|c|c|c|}
%\arrayrulecolor{blue}
 \hline
 \multirow{3}{*}{} & \multicolumn{2}{c|}{Idealistic} & 
    \multicolumn{2}{c|}{Frame-Based}\\
\cline{2-5}
  & {Slow Process} & {Fast Process} & {Slow Process}  & {Fast Process}\\
 \hline
  {\gls{tare}} &  {$0.068$} &  {$0.129$} & {$0.223$} & {$0.250$}\\ 
 \hline
{\gls{tacae}} &  {$0.204$} &  {$0.387$} & {$0.638$} & {$0.550$}\\
 \hline
 {\gls{udc}} &  {$0.614$} &  {$0.518$} & {$0.263$} & {$0.121$}\\
 \hline
\end{tabular}}
\vspace{-0.5cm}
    \label{tab:Slow_and_fast_DTMC}
\end{table}

\begin{figure}[!b]
\vspace{-0.5cm}
\centering 
\subfloat[]{
\includegraphics[height=4.2cm]{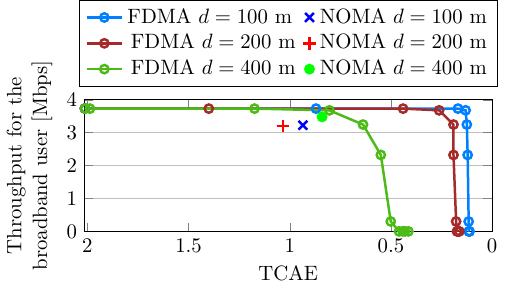}}\vspace{-0.5cm}\\
\subfloat[]{
\includegraphics[height=3.8cm]{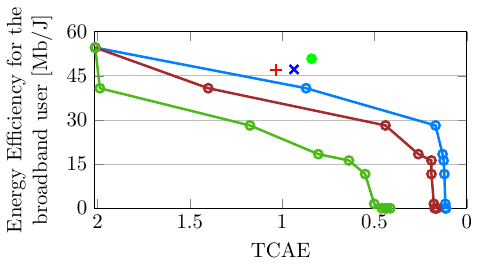}\vspace{-0.2cm}}
\caption{Throughput and Energy Efficiency performance of the broadband user versus the \gls{tacae} of the intermittent user.}
\label{fig:EEThroughputVsCAE}
\end{figure}
\par We now analyse the achievable trade-offs between the performance of the broadband user and the intermittent user in the Frame-Based model for the process $\left(p_s=0.1,\,q_s=0.15\right)$. {Fig.~\ref{fig:EEThroughputVsCAE}(a) illustrates the Pareto front vis-à-vis throughput performance of the broadband user and \gls{tacae} performance of the intermittent user for \gls{fdma} frames. The \gls{tacae} decreases by virtue of increasing $B_{2}$, hence the throughput performance of the broadband user decreases, as $B_{1}$ decreases. It is also evident that with increasing $d$, the area of the corresponding Pareto front decreases.} Nevertheless, it is interesting to observe from Fig.~\ref{fig:EEThroughputVsCAE}(a) that a marginal throughput reduction is needed to accommodate the transmissions of the intermittent user in \gls{fdma} and improve its error metrics. However, the throughput drops significantly after a \gls{tacae}, which is naturally different for different distances. This is due to the fact that maintaining throughput performance with decreasing $B_{1}$ comes at the expense of increasing transmission power by the broadband user. This is reflected in the energy efficiency performance illustrated in Fig.~\ref{fig:EEThroughputVsCAE}(b). A  consistent decline in the energy efficiency of the broadband user is observed with decreasing \gls{tacae} of the intermittent user. Ultimately, as $P_{\textrm{max}}$ becomes insufficient to achieve $r_{1}^{\textrm{max}}$; see equation \eqref{eq:Power_BBUser}; the throughput performance also starts decreasing; Fig.~\ref{fig:EEThroughputVsCAE}(a).
\par {Next, we focus on the performance of the two users for \gls{noma} frames, i.e., $B_{3}=B$. Fig.~\ref{fig:EEThroughputVsCAE}(a) and Fig.~\ref{fig:EEThroughputVsCAE}(b)  illustrate that with \gls{noma}, the broadband user achieves high throughput and energy efficiency, %performances, 
whereas the intermittent user achieves a modest \gls{tacae} performance. Interestingly, in Fig.~\ref{fig:EEThroughputVsCAE}(b), the performance metrics of both users with \gls{noma} lie outside of the \gls{fdma} Pareto front. Specifically, to reduce  \gls{tacae} by half, the drop in energy efficiency with \gls{noma} is significantly less than that with \gls{fdma}. Note that, the \gls{tacae} of the intermittent user being worst for $d=200$ is due to small differences between the \gls{snr}s of the users leading to low capture probabilities. Regarding \gls{udc}, Table \ref{tab:CAEVsSC} illustrates that   \gls{noma} achieves similar \gls{tacae} performance to \gls{fdma}, without any increase in the \gls{udc}. Therefore, \gls{noma} achieves a better performance trade-off between the two users.}
\begin{table}[!t]
\centering
\caption{Performance of the intermittent user: \gls{fdma} vs.  \gls{noma}.}\vspace{-0.15cm}
  {\begin{tabular}{|l|c|c|c|c|c|c|}
%\arrayrulecolor{blue}
 \hline
 \multirow{3}{*}{} & \multicolumn{2}{c|}{\gls{tacae}} & 
    \multicolumn{2}{c|}{\gls{udc}}\\
\cline{2-5}
  & {\gls{fdma}} & {\gls{noma}} & {\gls{fdma}}  & {\gls{noma}}\\
 \hline
  {$D=100$} &  {$0.92$} &  {$0.94$} & {$0.44$} & {$0.45$}\\ 
 \hline
{$D=200$} &  {$1.06$} &  {$1.03$} & {$0.54$} & {$0.52$}\\
 \hline
 {$D=400$} &  {$0.84$} &  {$0.84$} & {$0.39$} & {$0.38$}\\
 \hline
\end{tabular}}
\vspace{-0.5cm}
    \label{tab:CAEVsSC}
\end{table}
\section{Conclusion}
{This work investigated and analysed the task-dependent error-metrics performance of a real-time remote tracking and actuation scenario in a Frame-Based resource sharing scenario. A communication model predicated on `idealistic' assumptions in conventional studies on \textcolor{black}{real-time remote reconstruction} was considered as the baseline. We demonstrated that the error-metrics performance gain of the baseline model in conventional systems over Frame-Based model is achieved at the expense of a disproportionately high update-delivery cost. After that, for the Frame-Based model, we analysed the performance trade-off between the intermittent user effectuating the real-time remote tracking and a broadband user, for \gls{fdma} and \gls{noma} schemes. \textcolor{black}{We demonstrated that \gls{noma} achieves a better performance trade-off between the users, particularly when energy efficiency is the  measure for the broadband user.} Building on this work, future studies will investigate joint optimization of the sampling policy and the uplink transmit power to improve the performance metrics of both users.}
\bibliographystyle{IEEEtran}
\bibliography{bib}

\end{document}